\documentclass[9pt,twocolumn,twoside]{osajnl}
\usepackage{graphicx}
\usepackage{amsmath}
\usepackage{mathtools}
\usepackage{amsfonts}
\usepackage{amssymb}
\usepackage{authblk}

\usepackage{tikz}
\usetikzlibrary{arrows,backgrounds,snakes,patterns}
\usetikzlibrary{shapes,arrows,chains}
\usepackage{verbatim}
\usepackage{booktabs}

\usepackage[flushleft]{threeparttable}

\journal{ol} 

\setboolean{shortarticle}{true}

\title{Aperiodic phase masks for inscribing complex multi-notch OH-emission filters for astronomy}

\author[1,*]{Kalaga Madhav}
\author[1]{Ziyang Zhang}
\author[1]{Martin M Roth}

\affil[1]{innoFSPEC, Leibniz-Institut für Astrophysik Potsdam, An der Sternwarte 16, Potsdam, Germany, 14482}

\affil[*]{Corresponding author: kmadhav@aip.de}

\ociscodes{(060.3735) Fiber Bragg gratings; (050.5080) Phase shift; (060.2340) Fiber optics components; (120.2440) Filters; (230.1480) Bragg reflectors}

\begin{abstract}
We demonstrate for the first time, a new type of aperiodic phase mask (APM) for fabricating multi-channel aperiodic fiber Bragg gratings. The mask is made of individual diffraction phase gratings with discrete unequal phase-steps incorporated at periodic locations. The diffraction at the discrete phase-steps in the phase mask produces corresponding half phase-steps at periodic locations along the fiber.  The accumulated phase, along with index modulation, generates the desired multinotch reflection spectrum. Complex fiber Bragg grating filters fabricated using APM, in a standard phase mask based fabrication setup, can be used to simultaneously suppress multiple aperiodic OH emission wavelengths in near infrared (NIR) existing in upper atmosphere, and increase the sensitivity of ground based telescopes.
\end{abstract}

\setboolean{displaycopyright}{true}

\begin{document}

\maketitle

\section{Introduction}

Observation at near infrared wavelengths (NIR) between 0.9 to 2.5$\mu$m are critical for modern astrophysics, as they provide access to objects heavily obscured by dust extinction, e.g. the supermassive black hole at the galactic center, and star forming regions, to cool objects like AGB stars, to the high redshift universe, etc. -- to name but a few. Furthermore, the availability of high sensitivity large format image sensors and the advent of adaptive optics have made the NIR an extremely attractive wavelength range such that the new generation of large ground-based telescopes like the ELT, TMT, or GMT must be considered mainly NIR facilities. However, observations of faint objects in the NIR from the ground are overwhelmed by a sky background emission line spectrum that is typically 1000 times brighter than the NIR light from the objects of interest. The emission occurs due to de-excitation of atmospheric hydroxyl (OH) molecules in a cold layer of 6-10 km thickness at altitudes of 90km . At the central wavelengths of these emission lines, the signal of faint objects is heavily affected by the photon shot noise and strong residuals associated with the OH lines, so no reliable data can be recorded. Instead, one has to resort to the interline continuum, a technique also known as OH avoidance \cite{Martini}. However, even when resorting to observations between the bright OH lines, it has been discovered that the faint extended wings, that are due to scattered light occurring inevitably within the spectrograph, are still bright enough to affect the detection limit of faint objects in the continuum. Therefore, it has been considered to filter out the OH lines at high dispersion, however, first concepts have in reality not provided convincing results (e.g. \cite{Piche,Ennico,Maihara}). A radically new idea was proposed by \cite{Bland} which consists of a filter placed in front of the optical system before the light enters the spectrograph, thus giving nowhere an opportunity to create scattered light \cite{Ellis1,Trinh1}. In order to suppress or filter out the OH emission lines, an optical filter capable of $>$30dB suppression at the emission lines with bandwidths as small as 150nm and, with $>$0.5dB throughput outside and between the lines will be required. Fiber Bragg gratings (FBGs) are ideal candidates for filtering with the tight constraints and an aperiodic FBG (AFBG) filter capable of suppressing $\sim$100 lines has been previously demonstrated in the GNOSIS experiment \cite{Ellis2,Trinh2}. 

However, fabricating such filters, with good reproducibility is not a trivial task and requires accurate control of the intensity, phase and exposure length of a complex interference pattern over a moving photosensitive optical fiber. Simple or complex gratings can be fabricated through point-by-point (PbP) \cite{PbP} or line-by-line (LbL) \cite{LbL} inscription process using femtosecond lasers and de-phasing methods \cite{Buryak}. Ultra-long gratings have been fabricated using electro-optic modulators (EOMs) \cite{Raman} in push-pull configuration, and complex OH filters have been fabricated using acousto-optic modulators (AOMs) \cite{JAR}. E- or A-OMS fabrication techniques generate a \textquotedblleft running-interference\textquotedblright  pattern, similar to a rack-and-pinion, that is synchronised with the velocity of optical fiber, requiring precise control on the intensity, focus spot size, and velocity over a long length in real time. Phase mask offers a convenience that previously mentioned methods do not offer. By transferring the complexity of fabrication, such as in femtosecond, EOM or AOM techniques, to the one-time manufactured complexity in the phase mask, the convenience is preserved.  Requiring no moving parts, or stringent alignment, the complex phase mask can be used off-the-shelf in a standard UV based FBG inscription setup. In this paper, we introduce for the first time the design of an aperiodic phase mask to inscribe multi-channel aperiodic filters in hydrogenated or doped photosensitive fibers, in order to suppress the night-sky OH emission.

\section{Aperiodic Bragg grating}

The index modulation $\Delta n_g$ and phase $\phi_g$, of the complex grating can be reconstructed from the desired reflection spectrum $|r|$, by using layer peeling method described in \cite{Skaar1,Skaar2}. For the design of APM, we selected OH sky emission lines in H-band ranging from 1400nm to 1700nm \cite{Rousselot}.  For compatibility with future tests on PRAXIS \cite{praxis} system, that uses the existing GNOSIS filters, the full-width half maximum (FWHM), transmission, and wavelength of OH lines are selected as given in Tables.(1,2) in \cite{Trinh1}. The aperiodic filter is defined by \cite{Cao},

\begin{equation}\label{eq:desiredspectrum}
\begin{multlined}
|r\big(\lambda\big)| =\sqrt{R_i}\sum_{i=1}^{N}{\exp \Bigg[-\Bigg\{\frac{2\pi n_{eff}}{p_i}\Big(\frac{1}{\lambda}-\frac{1}{\lambda_i}\Big)\Bigg\}^{q_i}\Bigg]}\times \\
 \exp \Bigg[i2\pi n_{eff}\Bigg(\frac{1}{\lambda}-\frac{1}{\lambda_0}\Bigg)g_i\Bigg]
\end{multlined}
\end{equation}
where, $R_i$ is the desired reflectivity of individual OH-emission line filter, $N$ is the number of filters, $(p_i,q_i)$ define the shape of the individual filters, $n_{eff}$ is the effective index, $\lambda_0$ is the seed grating, which also defines the APM's pitch $\Lambda_m$, and $g_i$ is the individual channel's group delay. Since there exists an upper limit to the index change achievable in a fiber, $g_i$ can be optimised to reduce the maximum index modulation required. The right choice of $g_i$, or \textquotedblleft de-phasing\textquotedblright, \cite{Buryak,Cao} effectively spreads the individual gratings over the length of the grating, instead of crowding them in the same location spatially. We now describe the steps required to  design a mask that can be used in a standard UV inscription setup to fabricate a grating with a reflection spectrum defined by eq.(\ref{eq:desiredspectrum}).

\section{Design of aperiodic phase mask}
\label{sec:apm}

\begin{figure}[htbp]
\centering
\includegraphics[width=50mm]{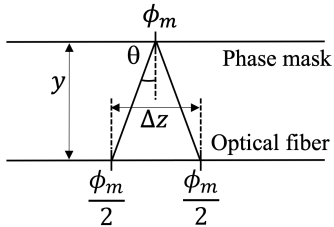}
\caption{Schematic showing the propagation of phase mask phase to fiber grating phase.}
\label{fig:mask_fiber1}
\end{figure}

It is well known \cite{Sheng1,Sheng2,Sheng3} that when a $\phi_m$-shifted phase mask is used in side-writing technique for inscribing FBGs, the phase shift in the mask is split into two half-phase shifts ($\frac{\phi_m}{2}$), separated by $\Delta z=2y\tan\theta$ in the fiber, as shown in Fig.\ref{fig:mask_fiber1}, where, $y$ is the distance between the mask and the fiber core, and $\theta$ is the angle of diffraction of $\pm 1$ order.

The cumulative phase of the light propagating through the fiber gives the desired phase $\phi_g$ in the grating. For example, if a $\pi$-shifted phase mask is used to fabricate the grating, along the grating length there will be two locations with $\phi_g=\pi/2$ phase. In the transmission spectrum, we would see the characteristic single narrow transmission window at the filter center. With increasing $\phi_m$, the narrow transmission window within the Bragg stopband shifts to longer wavelength \cite{Janos}. To achieve the desired phase $\phi_g$ in the grating, the width of the groove, or phase-step, $\delta_m$, in the phase mask is given by,
\begin{equation}
\delta_m=\frac{\Lambda_{m}}{4\pi}\Big(2\pi+\phi_g\Big)
\label{eq:maskgap}
\end{equation}
For example, if $\phi_g=-\pi$ for the standard $\pi-$shifted FBG, we will require a phase mask with $\delta_m=\frac{\Lambda_m}{4}$ at the center of the mask of length $L$. If we use this mask for fabrication, in the fiber the phase will be split, $\phi_g=\big(-\pi/2,-\pi/2\big)$. By tuning $\delta_m$, or equivalently $\phi_m$, we can inscribe a desired $\phi_g$ in the grating at selected locations. Also, when $\delta_m=\Lambda_m/2$, then we get $\phi_g=0$, and a uniform phase mask. Introducing multiple phase shifts using high precision PZT translation was proposed in \cite{dai} for fabricating periodic or non-periodic high-channel count FBG.

In order to design a mask that can generate a grating with the desired aperiodic reflection spectrum defined by eq.(\ref{eq:desiredspectrum}), we will require the grating's phase ($\phi_g$). We use layer peeling (LP) technique to first derive the grating's complex coupling coefficient, $\kappa$, from which we can extract $\phi_g$. By knowing $\phi_g$, we can then calculate the phase-steps $\delta_m$ of the APM using eq.(\ref{eq:maskgap}). 

\begin{figure}[htbp]
\centering
\includegraphics[width=\linewidth]{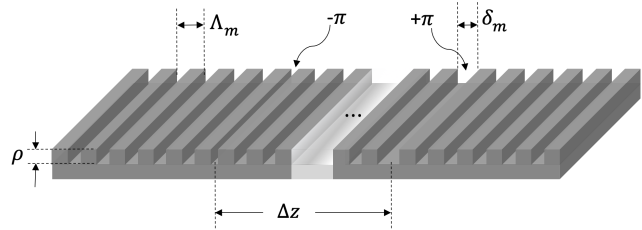}
\caption{Representative 2D model of a section of the APM showing two $\Lambda_m/4$ shifted grooves separated by $\Delta z$ corresponding to ($-\pi$, $+\pi$) phase. $\rho=\lambda_{uv}/2(n_{uv}-1)$ is the groove depth of the phase mask, defined by the wavelength $\lambda_{uv}$ of the laser used for fabrication, and the refractive index $n_{uv}$ of the mask material.}
\label{fig:apm}
\end{figure}

\section{Simulation and discussion}
\label{sec:simulation}
\begin{figure}[htbp]
\centering
\includegraphics[width=\linewidth]{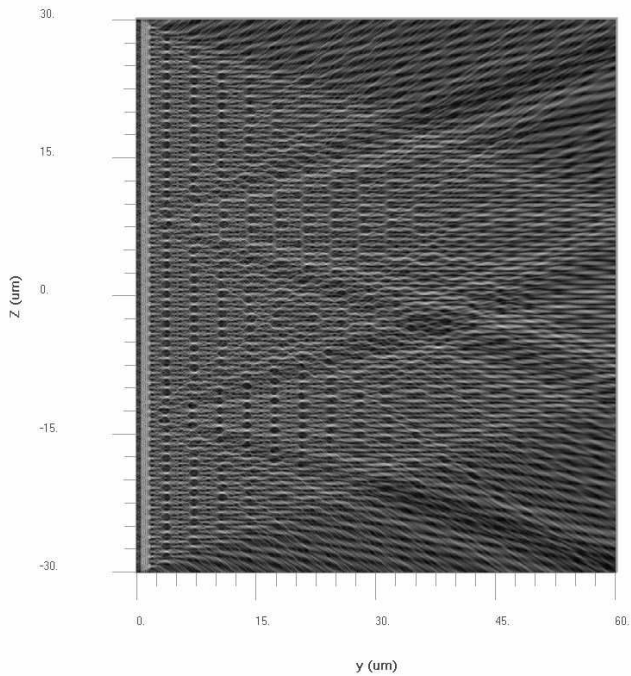}
\caption{Near field of phase mask with two phase steps of $-\pi$  and $+\pi$.}
\label{fig:maskfdtd}
\end{figure}

Fig.\ref{fig:maskfdtd} shows the FDTD simulation for phase steps at two locations on a phase mask of $60\mu m$ length, separated by 20$\mu m$, where mask period $\Lambda_m=1.064\mu m$. The two phase steps $\delta_m=\Big\{\frac{\Lambda_m}{4},\frac{3\Lambda_m}{4}\Big\}$ corresponding to mask phases $\phi_m=\big\{-\pi,+\pi\big\}$, respectively, result in four half-phase regions in the fiber.

\begin{figure}[!htbp]
\centering
\includegraphics[width=\linewidth]{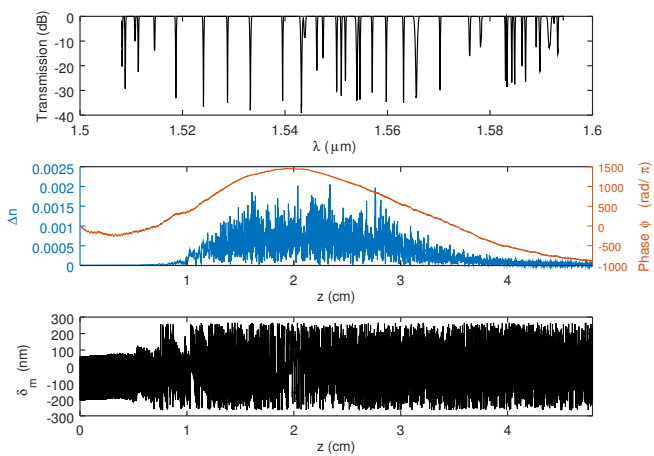}
\caption{Transmission spectrum, index modulation, phase and APM groove width for fabricating an aperiodic filter.}
\label{fig:layer}
\end{figure}

For designing the APM specifically made to fabricate an aperiodic grating, we first define the filter characteristics, such as, transmission, FWHM and channel center. We chose $N=37$ OH-emission lines between 1500nm to 1600nm, and $L=47.9$mm. Using LP, and the desired filter spectrum constructed using eq.(\ref{eq:desiredspectrum}), we find the index modulation $\Delta n$ and $\phi_g$, as shown in Fig.\ref{fig:layer}.  For covering the filter's bandwidth ($\beta$), we will require a $\phi_g$, or $\delta_m$ discretization interval or layer thickness in layer peeling \cite{Skaar2}, $\Delta z=\pi / \beta=9.58 \mu m$. We use eq.(\ref{eq:maskgap}) to calculate the APM's groove width $\delta_m$ from $\phi_g$. Since $\phi_g=[-\pi,+\pi]$ rad, we have $\delta_m=[266.32,798.96]$ nm.

$\delta_m$ can be incorporated in the mask as a nonlinear chirp, where the groove width $\delta_m$, varies continuously over the length of the mask. To achieve accurate continuously varying $\delta_m$ using e-beam process would be challenging. Alternatively, we design the mask with a global mask pitch of $\Lambda_m=1065.28$nm, corresponding to the seed grating $\lambda_0=1550$nm, and at discrete locations at intervals of $\Delta z$ along the seed phase mask, we incorporate grooves of width $\delta_m$ defined by eq.(\ref{eq:maskgap}). An example of a mask segment with two $\delta_m$ is shown in Fig.\ref{fig:apm}.

\section{Conclusion}
We have shown the steps involved in transferring the spatial structure of an aperiodic fiber Bragg grating to the corresponding structure in an aperiodic phase mask. Fabrication of the mask with the desired groove accuracy at periodic intervals, or as a continuous chirp is not a trivial task. However with recent advances in e-beam processes, the accuracy required to reproduce $\delta_m$ is a reality. Fabrication of APMs based on the method described in this paper is currently ongoing. With APMs, the complexity of alignment in fabrication setups such as EOM, AOM or femtosecond laser, is now transferred to the fabricated complexity in the mask, facilitating the use of standard phase mask fabrication to inscribe complex gratings.

\section*{Acknowledgements}

This work is supported by the BMBF project “Meta-ZiK Astrooptics” (grant no. 03Z22A511). 



\end{document}